\documentclass[12pt,worldsci]{article}
\usepackage{graphicx}
\begin{document}
\title{AN ANALYSIS OF 4-QUARK ENERGIES \\
IN SU(2) LATTICE MONTE CARLO}
\author{Sadataka Furui \\
{\em School of Sci. and Engr., Teikyo Univ., 
 Utsunomiya 320-8551, Japan}\\
%(e-mail furui@dream.ics.teikyo-u.ac.jp)
\vspace{0.3cm}
and \\
%\vspace{0.3cm}
Bilal Masud \\
{\em  Centre for High Energy Physics, Punjab Univ.,
      Lahore 54590, Pakistan }}
%(e-mail masud@chep-pu.edunet.sdnpk.undp.org)
\maketitle
\setlength{\baselineskip}{2.6ex}

\vspace{0.7cm}
\begin{abstract}
Energies of four-quark systems with the tetrahedral geometry measured by the 
static quenched SU(2) lattice monte carlo method are analyzed
by parametrizing the gluon overlap
factor in the form $exp(-[b_sE{\cal A}+\sqrt{b_s}F{\cal P}])$, where 
${\cal A}$ and ${\cal P}$ are the area and perimeter defined mainly 
by the positions of the four quarks, $b_s$ is the string constant in the 
2-quark potentials and $E,F$ are constants.   
\end{abstract}
\vspace{0.7cm}

\section{Introduction}

For the purpose of understanding meson-meson interactions from the QCD,
we studied the ground and the 1st excited state energy of 4-quark system
measured by the SU(2) lattice Monte-Carlo. The configuration of the four
quarks that we consider are large square(LS), rectangular(R), 
tilted rectangular(TR), linear(L), quadrilateral(Q), non-planar(NP) and
tetrahedral(TH).  

There are three possible choices of the colour singlet pairs which are denoted
 by A(14,23), B(12,34) and C(13,24). (We interchange
the definition of$A,B$ from\cite{GM}.) Identifying these bases as $P_1, P_2$ and $P_3$, we obtain
the eigen-energies by diagonalizing $3\times 3$ matrix of
\begin{equation}
W^T_{ij}=<P_i| \hat T^T|P_j>
\end{equation}
where $\hat T=exp(-a\hat H)$ is the transfer matrix.
Using a trial wave function $\psi=\sum_i a_i|P_i>$,
the eigenvalue equation
\begin{equation}
W^T_{ij}a_j^T=\lambda^{(T)}W_{ij}^{T-1}a_j^T
\end{equation}
is solved to get
$exp(-a V_2)=\lambda^{(T)}$ for large T\cite{GM}.

\section{The model with the gluon overlap factor}

We observed that among the three bases, one base can be treated as a linear
 combination of the other two, but within a base in the coarse lattice,
when the shortest path
between a quark and an antiquark is not along any link, there are several
possible configurations of links, which are specified by the parity.
 When there are m(n) parity eigenstates in A(B), we solve;
{\small
\[
det\left( \left[   
   \begin{array}{cccccc}
     V_{A} & \ldots & c_{A1m} & {V_{AB}f}_{11} & \ldots & {V_{AB}f}_{1n} \\
     \vdots   & \ddots      & \vdots & \vdots  &  \ddots     & \vdots \\     
     c_{Am1}   & \ldots & V_{A} & {V_{AB}f}_{m1} & \ldots &  {V_{AB}f}_{mn} \\
     {V_{AB}f^*}_{11} & \ldots & {V_{AB}f^*}_{1m} & V_{B} & \ldots & c_{B1n} \\
     \vdots  & \ddots       & \vdots & \vdots  &  \ddots     & \vdots \\
     {V_{AB}f^*}_{n1} & \ldots & {V_{AB}f^*}_{nm} & c_{Bn1} & \ldots  & V_{B} 
    \end{array}
\right]\right.
\]

\[
-E\left[\left.   
    \begin{array}{cccccc}
      1 &  \ldots & 0 & f_{11}/2 & \ldots & f_{1n}/2 \\
      \vdots   &  \ddots   & \vdots & \vdots  &  \ddots     & \vdots \\     
      0 & \ldots &  1 & f_{m1}/2 & \ldots & f_{mn}/2 \\
      f^*_{11}/2 & \ldots & f^*_{1m}/2 & 1 & 0 & 0 \\
      \vdots   & \ddots      & \vdots & \vdots  &  \ddots     & \vdots \\     
      f^*_{n1}/2 & \ldots & f^*_{nm}/2 & 0 & 0 &1 
     \end{array}
\right] \right) =0
\]
}
 
We introduce the gluon overlap factor in the form $f_{ij}=exp(-[b_s E{\cal A}+
\sqrt{b_s}F_{ij}{\cal P}_{ij}])$, where ${\cal A}$ and ${\cal P}$ are the area and the perimeter. We
assume that the length of the zig-zag perimeter has the fractal dimension 2,
 and so it is fixed from that of the coarse lattice,
use the string constant $b_s$ measured in the simulation of the 2-quark 
potentials and fit constants $E$ and $F_{ij}${\cite{FGM}}. 
The length ${\cal P}_{ij}$ depends on the base but
the minimal area ${\cal A}$ are chosen to be independent of the base and 
estimated by 
the analytical form derived in the regular surface approximation. (The analysis
of NP of{\cite{FGM}} is revised, where the area depended on the bases.)
The perimeter dependent terms contain lattice artefacts and the physical
quantities are obtained by subtracting the artefacts.

The parameter $E$ for the area term  and $A_0$ for the self energy in the 
Linear configuration are fixed in the previous model{\cite{FGM}}. In the NP 
case we fitted
three coefficients $F$ corresponding to the three types of the perimeter 
lengths and a mixing parameter between the parity eigenstates\cite{FGM}. 
The details of the revised fitting are in {\cite{FM}}.

In the case of TH, when the length of the links are all equal($r=d$), the ground state
energy is doubly degenerate.
 When they differ($r\ne d$) we evaluate minimal surface area
for A-C($regular_1$) and for A-B($regular_2$) and redefine among $B$ and $C$,
 the base $B$ such that it is connected to $A$ by the smaller area.
We solve the secular equation of m=4 and n=4. 
\begin{figure}[h]
\begin{center}
\hspace{2mm}
\includegraphics[scale=.25]{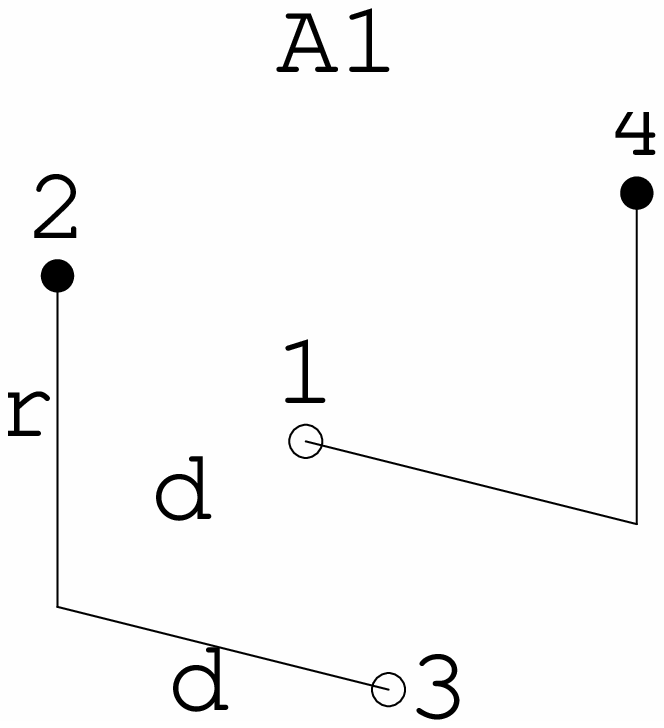}
\hspace{2mm}
\includegraphics[scale=.25]{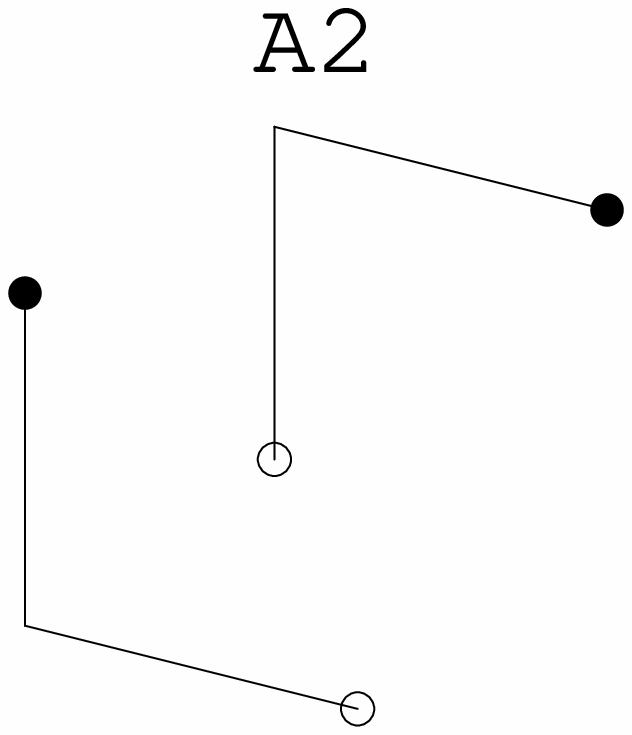}
\hspace{2mm}
\includegraphics[scale=.25]{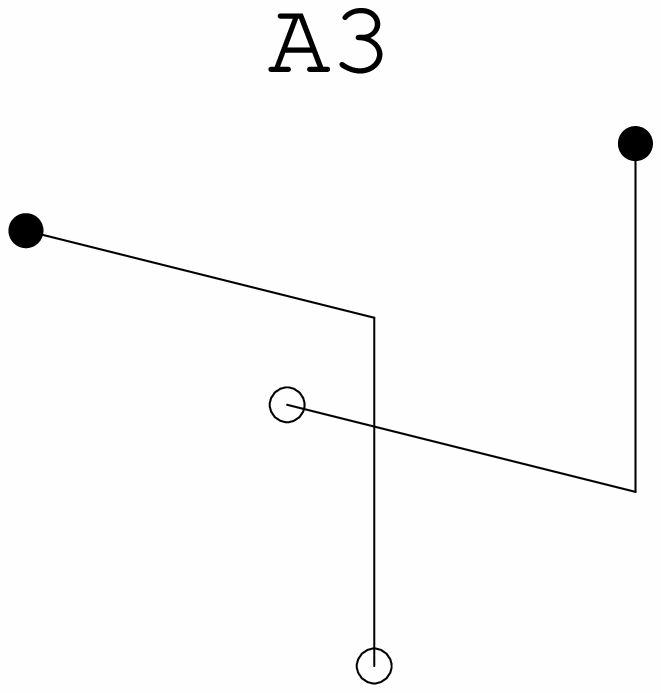}
\hspace{2mm}
\includegraphics[scale=.25]{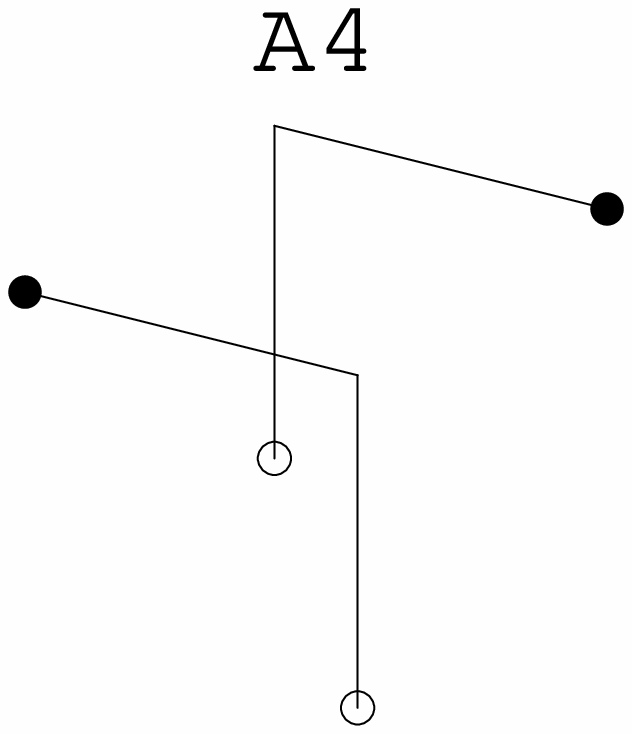}
\end{center}
\caption{The bases $A_1, A_2, A_3$ and $A_4$ in the TH configuration.}
\end{figure}

We explain the degeneracy of the ground state energy for the $r=d$ cases
and the ground and the first excited state eigen-energies of $r\ne d$ cases 
by introducing an additional coefficient $F$
corresponding to the new type of perimeter length
and a parameter that specifies the internal excitation of the two quark cluster.

\begin{table}[h]
\caption{Minimal surface area in regular surface approximation and the sum of two
triangles for the TH and the experimental and the fitted binding energy.}
\begin{center}
\begin{tabular}[h]{|r|r|r|r|r|r|r|r|}
\hline
%INSERT HERE THE DATA FILE.....START:
 & & & & & & & \\ r,d & $regular_1$& $regular_2$ & 2 triangles& $E_0$ &$E_0$(fit) & $E_1$ &$E_1$(fit)\\ \hline
        1   2   &   4.316 &  3.180 &      4.899 &-0.043 & -0.036 & 0.084 & 0.108 \\
        2   2   &   5.123 &  5.123 &      6.928 & -0.021 & -0.023 & -0.021 &-0.023\\
        3   2   &   6.209 &  7.260 &      9.381 & -0.008 & -0.011 &0.148 & 0.139\\
        2   3   &   10.22  &  8.528 &     12.369 &-0.040 & -0.043 & 0.051 & 0.072\\ 
        3   3   &   11.53  & 11.53  &      15.588& -0.028 & -0.027 & -0.028 & -0.027 \\
        4   3   &   13.11 &  14.71  &     19.209 & -0.007 & -0.012 & 0.110 & 0.115\\
        3   4   &   18.69 &  16.45  &     23.324 & -0.041 & -0.050 & -0.032 & -0.032\\
        4   4   &   20.49 &  20.49  &     27.713 & -0.030 & -0.031 & -0.030 & -0.031\\
        5   4   &   22.56 &  24.71  &     32.496 &-0.010 & -0.014 & 0.089 & 0.093\\
\hline
\end{tabular}
\end{center}
\end{table}

\section{Discussion and Conclusions}

In an analysis of 4-quark energy of the tetrahedral geometry,  
Green and Pennanen{\cite{GP}} proposed a 6-basis model in which the ground
($A_{1g}A_{1g}$) and an excited state ($E_uE_u$) are considered for each of
 the configuration A, B and C. Our choice corresponds to  
$A_{1g}A_{1g}, E_uE_u, A_{1g}E_u$ and $E_uA_{1g}$ for each of the
configuration A and B. Our estimation of the minimal surface is more accurate
than their triangular approximation.
The relatively large coefficients of ${\cal P}$ suggests that
the lattice artefact is still large. 
 
\end{document}